\documentclass{article}
\usepackage[margin=3cm]{geometry}

\usepackage{amsmath,amssymb,amsfonts}
\usepackage{mathrsfs} 
\usepackage{mathtools}
\usepackage{orcidlink}
\usepackage{subcaption}
\usepackage{float}

\usepackage{graphicx,color,epsfig}
\usepackage[nocompress]{cite}

\usepackage{authblk}

\usepackage{enumitem}

\begin{document}

\renewcommand{\arraystretch}{1.3}
\setlength{\tabcolsep}{1ex}

\title{Bulk-boundary decompositions and exact perfect fluid stellar distributions   }
\author[1]{Sudan Hansraj\orcidlink{0000-0002-8305-7015}\footnote{Email: hansrajs@ukzn.ac.za}}
\author[2,1]{Christian G B\"ohmer \orcidlink{0000-0002-9066-5967}\footnote{Email: c.boehmer@ucl.ac.uk }}
\author[1]{Ndumiso Buthelezi\orcidlink{0000-0001-8825-9903}\footnote{Email:  215033901@stu.ukzn.ac.za}}
\affil[1]{Astrophysics Research Centre, School of Mathematics, Statistics \authorcr and Computer Science, University of KwaZulu-Natal, \authorcr Private Bag X54001, Durban 4000, South Africa\medskip}
\affil[2]{Department of Mathematics, University College London, \authorcr 
Gower Street, London WC1E 6BT, UK}

\date{8 Oct 2026}

\maketitle

\begin{abstract}
The decomposition of the Ricci scalar in terms of bulk and boundary terms gives rise to an interesting  gravitational field theory as proposed by B\"ohmer and Jensko (BJ). BJ went on to suggest functional forms of the bulk contributions in the Lagrangian. For the purposes of astrophysical modelling the boundary terms are non-dynamical and the entire theory is built from the bulk contribution only. The field equations for a stellar perfect fluid distribution are obtained against the background of a conformostatic metric. Since the linear form of the function of the bulk term is equivalent to Einstein we probe the next level of complexity namely the pure quadratic case. The field equations are substantially more complicated than Einstein's equations. The equation of pressure isotropy proves intractable even with simple choices of one of the geometric potential functions. Accordingly we analyse the case of the potentials varying inversely with the radius. Viable models do not arise in this case and so we postulate a proportional relationship between the potentials. It turns out that a physically reasonable shell of isotropic fluid may be constructed that satisfies elementary physical requirements. Finally we investigate the isothermal fluid where both density and pressure go as the inverse square law of the radius. This results in fluid behaviour that is comparable to that in the corresponding Einstein model. 
\end{abstract}

\section{Introduction}\label{intro:Sec}

The standard theory of gravitation, Einstein’s general relativity, has inspired numerous extensions and modifications in recent years. This trend is driven by several theoretical and observational challenges: the theory's inability to explain the late‑time accelerated expansion of the universe without invoking exotic dark energy, its untested nature on cosmological length scales and in extreme‑gravity regimes, and its non‑renormalizability within the framework of quantum field theory \cite{Goroff1985,tHooft1974}.

A prominent class of modifications is Lovelock gravity \cite{love1,love2}, which preserves desirable features such as diffeomorphism invariance, consistency with the Bianchi identities, second‑order equations of motion, and Ostrogradsky stability. Its second‑order polynomial yields Einstein‑Gauss‑Bonnet (EGB) gravity, a theory supported by its emergence in the low‑energy effective action of string theory. A potential limitation is that higher‑curvature Lovelock terms become dynamical only in spacetime dimensions greater than four. Recent proposals to rescale the coupling constant to activate the Gauss‑Bonnet term in four dimensions \cite{Glavan} have been controversial, though the approach is considered safe under highly symmetric geometries like spherical symmetry \cite{GursesSismaniTekin}. A natural extension within this framework is `$f(G)$ gravity,' where the action is a general function of the Gauss‑Bonnet invariant $G$. In this work, we focus on a related theory proposed by B\"ohmer and Jensko \cite{BJ}, which we refer to as $f(G, B)$ gravity (though the boundary term $B$ itself will not be central to our analysis).

Other notable alternatives include $f(R)$ gravity, which, while cosmologically viable, introduces fourth‑order field equations and associated ghost instabilities \cite{Nojiri2007,Sotiriou2010,DeFelice2010}. Theories such as $f(R, T)$ \cite{Harko2011}, Rastall gravity \cite{Rastall1972}, and unimodular gravity \cite{HansrajEllis1} are geometrically akin to general relativity but typically violate the standard conservation of energy‑momentum. Meanwhile, `$f(Q)$ gravity,’ based on non‑metricity, appears promising only in its linear form, where it reproduces general relativity. While many of these modified theories successfully model cosmic acceleration, they often possess significant weaknesses, particularly in astrophysical contexts. For instance, higher‑order $f(R)$ and $f(G)$ theories frequently struggle to produce physically viable compact star models for perfect fluids. Constructing such models typically requires an artificial identification of higher‑order geometric terms with matter content, leading to unfamiliar fluid distributions. Consequently, obtaining models with a realistic equation of state remains a considerable challenge.

In a unifying approach, B\"ohmer and Jensko \cite{BJ} explored the role of boundary terms in the decomposition of the Ricci scalar within metric‑affine geometries. They noted that formulations like the Teleparallel Equivalent of General Relativity (TEGR) \cite{Maluf2013} and the Einstein‑Hilbert action itself are Lorentz invariant only up to a boundary term. While the well‑studied $f(R)$ theory \cite{Starobinsky} maintains full invariance, it suffers from higher‑order equations. B\"ohmer and Jensko argued that by separating the Ricci scalar into bulk and boundary contributions and constructing an action from the bulk term alone, a theory with second‑order field equations could be achieved. Their investigation of boundary terms revealed that $f(R)$ gravity emerges as the unique Lorentz‑invariant theory, while $f(T)$ gravity (based on torsion) is the unique second‑order theory. This framework identifies three distinct boundary terms whose limits recover $f(T)$, $f(Q)$, and $f(R)$ theories \cite{BJ}. A subsequent extension within the Palatini formalism further clarified the role of torsion and non‑metricity for nonlinear boundary terms \cite{BJ2}.

This seminal work has stimulated further research into the connections between modified gravity and boundary terms. For example, Capozziello et al. \cite{Capozziello2023} examined the boundary term in $f(Q, B)$ gravity and its contrast with $f(Q)$ theory in the limit $B \to 0$. Shabani et al. \cite{Shabani2024} and Kadam et al. \cite{Kadam2023} employed dynamical systems techniques to study the cosmological implications of boundary terms in symmetric teleparallel gravity and extended teleparallel gravity, respectively. Similarly, Vishwakarma and Shah \cite{Vishwakarma2024} analyzed higher‑curvature gravity models incorporating specific boundary‑term functions.

Beyond the boundary-term programme, the broader landscape of stellar structure in modified gravity provides important context for the present study. Compact-star models have been examined extensively in $f(R)$ gravity, where hydrostatic equilibrium and realistic neutron-star sequences have been constructed despite the fourth-order field equations \cite{Capozziello2011s,Astashenok2013,Yazadjiev2014}. Parallel efforts within the symmetric-teleparallel family have produced static spherically symmetric solutions in $f(Q)$ gravity, in which the affine connection is promoted to a dynamical field and can support genuine beyond-GR configurations \cite{DAmbrosio2021}. The recently established correspondence between scale-dependent gravity and scalar-tensor theory \cite{Neckam2025}, together with the running-coupling viewpoint in which Newton's constant itself acquires scale dependence \cite{Koch2022}, further illustrates how modifications of geometric or quantum-gravitational origin reshape self-gravitating matter. Quasi-normal and greybody analyses of $f(R)$ black holes \cite{Panotopoulos2026} probe the observational fingerprints of higher-curvature corrections along a complementary line. The bulk-boundary decomposition studied here isolates the purely first-derivative bulk sector and asks whether it, on its own, can sustain physically acceptable perfect-fluid interiors.

Within this landscape, isotropic stellar models—where the radial and tangential pressures are equal—provide a foundational framework for studying compact objects. The isotropic condition simplifies the matter description to a perfect fluid, though it introduces a constraint that complicates the solution of the field equations. Nevertheless, numerous exact isotropic solutions are known in spherical symmetry \cite{finchskea}. These models are theoretically vital for understanding the fundamental physics of relativistic stars and serve as benchmarks for more complex, anisotropic scenarios. Isotropic solutions have been extensively studied in general relativity \cite{Tolman1939, Buchdahl1959} and in various modified gravity settings, including Gauss‑Bonnet‑inspired theories \cite{N2Dadhich, Hansrajt1, Ivanov2002}.

 Exact solutions of the gravitational field equations are very important because they allow us to understand the behaviour of a model and reveal all aspects of the spacetime geometry as well as the physics.  A useful guide to physically relevant solutions is contained in Stephani et al \cite{Stephani}. In contrast numerical models, can give good estimates of physical variables such as density and pressure but cannot show the exact metric functions everywhere. Exact models therefore help us see how all the variables behave throughout the star and serve as valuable checks on numerical and approximate methods. However, finding exact solutions requires simplifying assumptions, such as prescribing one of the unknown functions and then integrating the field equations, because the field equations are highly nonlinear and cannot be solved in general. This approach was used successfully by Tolman \cite{Tolman1939} who wrote the master field equation in a mathematically expedient way to extract exact solutions. Eight classes of solutions were reported: some contained known solutions while some new models such as Tolman IV were shown to be physically viable. Therefore this technique should not be dismissed.  However it must be conceded that with this approach, that is somewhat {\it{ad hoc}}, it is understandable that many exact models fail some elementary physical tests; for example, they may not admit a realistic equation of state. In the literature, there are over 120 known static, spherically symmetric perfect fluid exact solutions to the Einstein field equations for the perfect fluid star. A detailed list of static solutions that satisfy strict physical conditions for spherically symmetric perfect fluid models was compiled by Delgaty and Lake  as well as Finch and Skea \cite{finchskea,delgaty}. One can observe from this list that the number of solutions that are physically applicable are fewer than 10. This illustrates the complexity of modelling stars through the complicated system of differential equations.

 The solution generating process follows a pattern. For an isotropic stellar model, the main tool we use is the isotropy equation, where the radial pressure $p_r$ equals the tangential pressure $p_t$. By setting $p_r = p_t$, the field equations reduce to a single differential equation, which can then be solved to find the interior geometry of the star \cite{finchskea,makharko}. To solve this equation, we usually pick a simple form (ansatz) for one of the metric functions, substitute it into the pressure isotropy condition and then solve for the remaining unknown function. Over the years, several methods have been developed to integrate this equation in general relativity. For example, the isotropy condition can be transformed into a Riccati equation, which in some cases can be solved analytically as shown by Hajj-Boutros \cite{hajjboutros}. Other approaches include choosing a specific density or a particular form of one of the metric potentials, which reduces the problem to an integrable form. Using these methods, researchers have generated a wide variety of exact solutions for isotropic and physically acceptable stellar models. In the present problem we have found through detailed empirical testing that the methods above do not lead to success.  We  need new insight to locate exact solutions.

 In a past work \cite{hans-cgb-nb}  we showed how the isotropy equation could be reduced to an autonomous equation and then the stability of solution trajectories could be studied. That approach suggested investigating forms for the metric potentials that vary inversely with the radius. After analysing some other simpler cases like constant potentials we endeavour to construct models with such an ansatz in the present article. Thereafter we consider the case of the potentials being related linearly. Finally we examine the isothermal property requiring an inverse square law fall off of the density and pressure. Such models have a baked-in equation of state however they also inherit the undesirable central singularity which must be dealt with.

\section{The $f(G, B)$ Gravity Formalism}

The $f(G, B)$ gravity theory, recently revitalized by B\"ohmer and Jensko \cite{BJ}, stems from a fundamental re-examination of the Einstein-Hilbert action. While standard General Relativity (GR) treats the Ricci scalar $R$ as an irreducible geometric quantity, the BJ approach exploits the fact that $R$ can be decomposed into a dynamic ``bulk'' component and a non-dynamical ``boundary'' component. 

\subsection{The Bulk-Boundary Decomposition}
Historically, boundary terms in the gravitational action were often discarded or countered by the Gibbons-Hawking-York term to ensure a well-defined variational problem. In the BJ formalism, this split is made explicit:
\begin{equation}
    R = G + B
\end{equation}
where $G$ represents the \textbf{bulk} (the part of the curvature containing only first derivatives of the metric) and $B$ represents the \textbf{boundary} (the total divergence term containing second derivatives). The Einstein-Hilbert action is thus recast as:
\begin{equation}
    S_{\rm EH} = \frac{1}{2 \kappa} \int (G + B) \sqrt{-g} \, d^4x
\end{equation}

The bulk term $G$ is defined by the contraction of Christoffel symbols:
\begin{equation}
    G = g^{\mu \nu}\left( \Gamma^{\lambda}_{\mu \sigma} \Gamma^{\sigma}_{\lambda \nu} - \Gamma^{\sigma}_{\mu \nu} \Gamma^{\lambda}_{\lambda \sigma} \right)
\end{equation}
Conversely, the boundary term $B$ is a pure divergence, expressed through the vector $B^\sigma$:
\begin{equation}
    B = \nabla_{\sigma} B^{\sigma} = \frac{1}{\sqrt{-g}} \partial_{\sigma} (\sqrt{-g} B^{\sigma})
\end{equation}
with the boundary vector defined as:
\begin{equation}
    B^{\sigma} = g^{\mu \nu} \Gamma^{\sigma}_{\mu \nu} - g^{\sigma \nu} \Gamma^{\lambda}_{\lambda \nu}
\end{equation}
where the Greek letters run from 1 to 4. 
The novelty in $f(G, B)$ gravity lies in treating $G$ and $B$ as independent arguments of a functional. Unlike $f(R)$ gravity, which forces a rigid coupling between bulk and boundary, $f(G, B)$ allows for a ``decoupled'' modification. This provides a mechanism to explore how the information encoded on the boundary of a spacetime region affects the interior dynamics, particularly in high-energy astrophysical environments where traditional GR might reach its limits.

\section{Field Equations of $f(G, B)$ Gravity}

To investigate stellar structures, we employ a static, conformally flat spatial manifold. We adopt the Cartesian-isotropic line element:
\begin{equation}
ds^2 = -e^{\nu(r)}dt^2 + e^{\mu(r)}\left(dx^2 + dy^2 + dz^2\right) \label{metric_cart}
\end{equation}
where $r = \sqrt{x^2 + y^2 + z^2}$. By transforming to spherical-isotropic coordinates $(r, \theta, \phi)$ via the standard identities:
\begin{equation}
x=r\sin\theta\cos\phi,\qquad y=r\sin\theta\sin\phi,\qquad z=r\cos\theta
\end{equation}
the metric is expressed as:
\begin{equation}
ds^{2} = -e^{\nu(r)}dt^{2} + e^{\mu(r)}\left(dr^{2} + r^{2}d\theta^{2} + r^{2}\sin^{2}\theta d\phi^{2}\right) \label{metric_sph}
\end{equation}

For the metric (\ref{metric_sph}), the bulk term $G$ simplifies to a compact form involving the radial derivatives:
\begin{equation} 
G = \frac{1}{2} e^{-\mu} \mu' \left(\mu' + 2\nu'\right) 
\end{equation}

\subsection{The Modified Dynamics}
Varying the $f(G, B)$ action with respect to the metric leads to the field equations. For a general $f(G)$ model, the equations for density $\rho$, radial pressure $p_r$, and tangential pressure $p_t$ are:

\begin{align}
\kappa \rho &= \frac{f}{2} - \frac{1}{2}e^{-\mu} f_G \left( 2\mu'' + \mu'\nu' + \mu'^2 + \frac{4\mu'}{r} \right) \nonumber \\ 
& \quad - \frac{1}{2}e^{-2\mu} f_{GG}\mu' \left( 2\mu'\nu'' - 2\mu'^2\nu' - \mu'^3 + 2\mu''\nu' + 2\mu'\mu'' \right) \label{eq_rho} \\
\kappa p_r &= \frac{e^{-\mu}}{2r} f_G \left( 2\mu' + r\mu'^2 + 2\nu' + 2r\nu'\mu' \right) - \frac{f}{2} \label{eq_pr} \\
\kappa p_t &= -\frac{f}{2} - \frac{1}{4} \left[ e^{-2\mu} f_{GG} (\mu'+\nu') \mathcal{K} + \frac{e^{-\mu} f_G}{r} \mathcal{L} \right] \label{eq_pt}
\end{align}
where $\mathcal{K} = -2 \mu '' \nu '+2 \mu '^2 \nu '+\mu '^3-2 \mu ' (\mu ''+\nu '')$ and $\mathcal{L} = r (2 (\mu ''+\nu '')+\nu '^2)+2 \mu ' (r \nu '+1)+r \mu '^2+2 \nu '$.

The pressure isotropy condition ($p_r = p_t$) results in:
\begin{multline}
  r f_{GG} (\mu'+\nu') \left[-2\mu''\nu' + 2\mu'^2\nu' + \mu'^3 - 2\mu'(\mu''+\nu'')\right] \nonumber \\ 
  + e^{\mu} f_G \left[-2r(\mu''+\nu'') + 2\mu'(r\nu'+1) + r\mu'^2 - r\nu'^2 + 2\nu'\right] = 0 \label{isotropy_mod}
\end{multline}
where it is notable that $f$ is absent save for its first and second derivatives. 
In the limit where $f(G) = G$, we recover the standard Einstein field equations. Conservation of the energy-momentum tensor $T^{ab}_{\,\,\,\,;b} = 0$ yields the hydrodynamic stability condition:
\begin{equation}
p' = -(\rho + p)\frac{\nu'}{2}
\end{equation}
for  isotropic pressures. 
Observe that this relationship remains functionally identical to GR, indicating that the modified geometry $f(G)$ acts as a source term while the local conservation laws remain intact.

\section{The pure Quadratic Formalism: $f(G) =  \epsilon G^2$}

Since the linear limit of $f(G)$ recovers General Relativity, the next logical extension in complexity is the pure quadratic form $f(G) =  \epsilon G^2$. Using the metric \eqref{metric_sph}, the gravitational field equations are expressed as:
\begin{align}
8 \kappa r\rho &= e^{-2 \mu } \epsilon \mu ' \left[ -32 r \mu '' \nu ' + 8 \mu '^2 (r \nu '-2) + 5 r \mu '^3 - 4 \mu ' (6 r \mu '' + 4 r \nu '' + r \nu '^2 + 8 \nu ') \right]  \label{106a} \\[1ex]
8 \kappa r p_r &= e^{-2 \mu } \left[  \epsilon \mu ' (\mu ' + 2 \nu ') (\mu ' (3 r \mu ' + 6 r \nu ' + 8) + 8 \nu ') \right] \label{106b}
\end{align}
and for the tangential pressure
\begin{multline}
8 \kappa r p_t = e^{-2 \mu }  \epsilon \bigl[  8 r \mu '' \nu '^2 + \mu '^3 (4 - 8 r \nu ') - 3 r \mu '^4 + 2 \mu '^2 (6 r (\mu '' + \nu '') \\- r \nu '^2 + 6 \nu ') + 4 \mu ' \nu ' (6 r \mu '' + 4 r \nu '' + r \nu '^2 + 2 \nu ') \bigr] \label{106c}
\end{multline}
where the  parameter $\epsilon$ is a real constant that may be set to unity with no loss of generality. 
 In this regime,  the condition for pressure isotropy ($p_r = p_t$) yields the master equation for the geometry:
\begin{align}
4 r \mu '' \nu '^2 - 2 \mu '^3 (5 r \nu ' + 1) - 3 r \mu '^4 &+ \mu '^2 (6 r (\mu '' + \nu '') - 7 r \nu '^2 - 6 \nu ') \nonumber \\ 
&+ 2 \mu ' \nu ' (6 r\mu '' + 4 r \nu '' + r \nu '^2 - 2 \nu ') = 0 \label{107d}
\end{align}
Any valid stellar model in pure $G^2$ gravity must satisfy this differential constraint. Clearly this equation is seriously more complicated than its Einstein counterpart and as expected even the simplest ansatze for one of the variables does not lead to exact solutions. Accordingly we will look at other approaches to locate exact models and then test their physical viability. 

\section{Vacuum Branches and Singularities}

The vanishing of the energy momentum tensor admits two families of solutions \cite{hans-cgb-nb} one of which is Minkowskian. The second branch generates  the  nontrivial metric for a curved vacuum in the form
\begin{equation}
ds^2 = -C(r + A)^4 dt^2 + E(r + A)^{-2} (dr^2 + r^2 d\Omega^2) \label{vac2}
\end{equation}
for a constant $A$. The singularity at $-A$ has been dealt within \cite{hans-cgb-nb}.  For stellar modeling, this vacuum must be matched to an interior fluid solution at a boundary $r = r_b$ where $p(r_b) = 0$, ensuring the singularity remains outside the physical region of interest.

\section{Constant potentials} 

In order to make some progress in finding exact solutions, we consider, at the outset,  the simplest possibility that of constant potentials. This approach was followed historically by Einstein also and is regarded as 'plucking the low hanging fruits' proverbially speaking. 

\subsection{Constant spatial potential}

In Einstein gravity, a constant space potential generates isothermal fluid spheres with  both density and pressures displaying an inverse square law fall-off with the radius \cite{Dadhich, Hansraj}. In the present case taking  $\mu = H$ a constant we find that  the pressure isotropy (\ref{107d}) is satisfied identically and from (\ref{106a}) the density vanishes. This suggests that we are dealing with a vacuum and any potential $\nu$ may be chosen. The pressure however remains non-zero and depends on $\nu$. This is an unacceptable physical situation and must consequently be considered defective. 

\subsection{ A constant temporally directed potential} 

When Einstein set the temporally directed potential to a constant in general relativity an exact solution was found where the density and pressure were both constant and the solution could not represent an isolated compact star \cite{Einstein1917}. This is known as the Einstein defective universe since these conclusions do not correspond to what is observed in the universe. In our case we let  $\nu = K$, a constant, in the pressure  isotropy condition 
(\ref{107d}) 
 which reduces to 
\begin{align}
   \mu'^2 \left( 2 \mu' + 3 r \mu'^2 - 6 r \mu'' \right) = 0 \label{421}
\end{align}

 for the constant potential. Ignoring the case $\mu'=0$ which has been considered elsewhere in this work equation (\ref{421}) may be  solved in the form 
\begin{align}
\mu =\log \left(\frac{c}{\left(r^{4/3}-d\right)^2}\right)  \label{109}
\end{align}
where $c$ and $d$ are integration constants. We now have the model completely determined. 

We find that the density and pressure are given by
\begin{align}
  \rho &= -\frac{512 \epsilon  \left(r^{4/3}-d\right)
    \left(2 r^{4/3}-9 d\right)}{27 c^2 \kappa  \sqrt[3]{r}}  \\
  p &= \frac{512 d \epsilon  \left(d-r^{4/3}\right)}{9 c^2 \kappa  \sqrt[3]{r}}
\end{align}
while the sound speed is given by
\begin{align}
  \frac{dp}{d\rho} &=
  -\frac{3 d \left(d+3 r^{4/3}\right)}{9 d^2+33 d r^{4/3}-14 r^{8/3}}.
\end{align}
Here are expressions for the energy conditions:
\begin{align}
  \rho - p &= -\frac{1024 \epsilon  \left(r^{4/3}-6 d\right) \left(r^{4/3}-d\right)}{27 c^2 \kappa  \sqrt[3]{r}} \\ 
  \rho + p &= -\frac{1024 \epsilon  \left(r^{4/3}-3 d\right) \left(r^{4/3}-d\right)}{27 c^2 \kappa  \sqrt[3]{r}} \\ 
\rho + 3p &= \frac{1024 r \epsilon  \left(d-r^{4/3}\right)}{27 c^2 \kappa }
\end{align}
and finally the adiabatic index of Chandrasekhar $\Gamma$ takes the form
\begin{align}
  \left(\frac{\rho +p}{p} \right)\frac{dp}{d\rho} &= \frac{2 \left(r^{4/3}-3 d\right) \left(d+3 r^{4/3}\right)}{-9 d^2-33 d r^{4/3}+14 r^{8/3}}.
\end{align}

We constructed plots for the model for the density, pressure, sound speed and adiabatic stability using a wide range of free parameter values.  We then concluded that the model is not physically viable. The density $\rho$ becomes negative, which immediately rules out the model as a description of a realistic star. The pressure $p$ remains positive and indeed vanishes for a finite radius. The sound speed $\frac{dp}{d\rho}$ is obviously infinite so cannot obey $0 \leq \frac{dp}{d\rho} \leq 1$ and the adiabatic stability $\left(\frac{\rho + p}{p}\right)\frac{dp}{d\rho}$ also fails for the same reason. Furthermore, the energy conditions expected to remain non-negative for realistic star are also violated for wide ranges of the parameters we studied. 
Overall, these problems show that models with a constant potential do not work in this approach. We conclude that this assumption does not lead to any realistic star model. Note that the interior Schwarzschild metric for an incompressible isotropic star in general relativity is also not physically reasonable since the sound-speed will be infinite. 

\section{Potentials varying inversely with radius}

In a previous work \cite{hans-cgb-nb} we have shown that the isotropy equation may be written in autonomous form thus facilitating a dynamical systems approach to the stability of the solution curves. That analysis also suggested functional forms for the potentials varying inversely with the radius. Let us now fully discuss this option.  We put $u = \mu'$ and $v = \nu'$  into equation (\ref{107d}) and rearrange in the form 
\begin{align}
Au' + Bv' + C = 0 
\end{align}
where $A = 2r(3u^2 + 6uv + 2v^2)$, $B = 2ru(3u + 4v)$ and $C = u(-3u^3 r - 10ru^2 v - 2u^2 - 7ruv^2 - 6uv + 2rv^3 - 4v^2$. 
It may also be observed that the equation remains invariant under the transformations $u \to u + c_1$ and $v \to v + c_2$ for arbitrary constants $c_1$ and $c_2$. It also admits a radial scaling so that we may get solution curves of the form
\begin{align}
u = \frac{a}{r} \qquad v = \frac{b}{r} 
\end{align}
that is 
\begin{align}
  \mu = a\ln r + k_1 \qquad \nu = b\ln r +k_2 
\end{align}
in the original variables and where $k_1$ and $k_2$ are constants.  
Plugging these into (\ref{107d}) will establish a relationship between $a$ and $b$: 
\begin{align}
a(a+2b)(3a^2 + 4ab + 8a - b^2 + 8b) = 0.
\end{align}
From this we get solutions $a = 0$ that is $\mu$ being constant. The case $a+2b = 0$ gives $\nu' = -\frac{1}{2} \mu'$. Both these cases have been considered before in the section $G = 0$ in \cite{hans-cgb-nb}. The last option is the quadratic curve 
\begin{align}
b = 2a+4 \pm \sqrt{7a^2 + 24a + 16}  \label{ab}
\end{align}
relating $a$ and $b$. This interplay between the values of $a$ and $b$ may be depicted in the plot Fig 1.
\begin{figure}[!htb]
    \centering
    \includegraphics[width=0.45\linewidth]{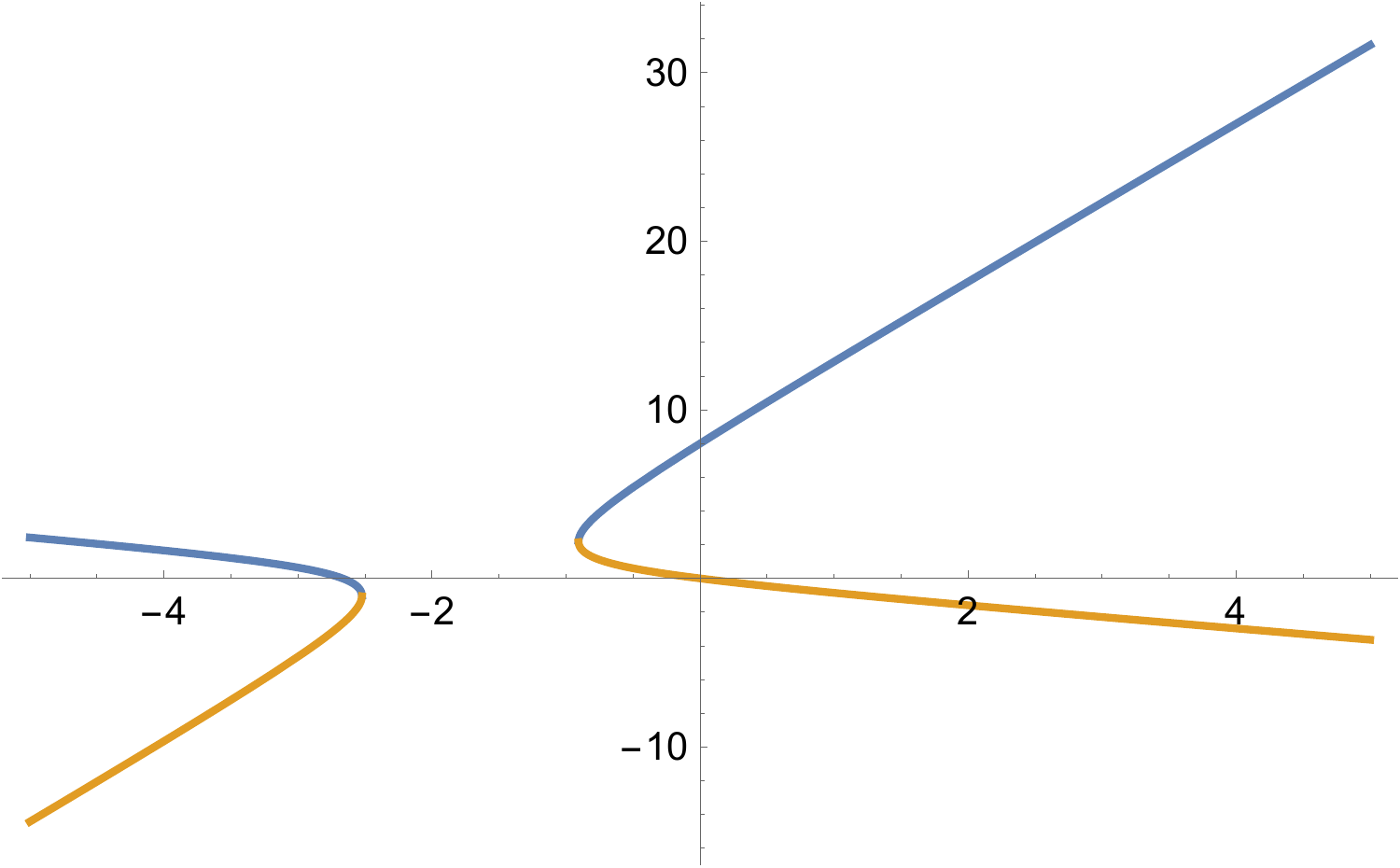}
    \caption{Feasible $a$ and $b$ values}
    \label{fig:placeholder}
\end{figure}
where the darker colour refers to the positive branch of (\ref{ab}). 
To construct realistic  models it will be useful to determine suitable $a$ and $b$ values such the physical constraints can be met. Integer values would be highly prized although not vital.  Considering the linear Diophantine equation $7a^2 + 24a+16 = w^2$ for $w$ an integer we detect that $a = 0$ ($\mu$ a constant) and $a = 24$ generate perfect squares. For $a = 24$, $b = 120$ or $-16$ giving solution families
\begin{align}
\mu = 24 \ln r + k_1, \, \, \nu  = 120 \ln r + k_2 \,\, {\mbox{or}} \,\, \nu = -16 \ln r + k_2
\end{align}
It can be checked that the next $a$ value that has integral $b$ value is $a = 480$. But it will not be instructive to inspect more cases. 

For now we keep the values of $a$ and $b$ general. A list of the dynamical quantities for this model will be listed below. The expressions for density and pressure are as follows
\begin{align}
  \rho &= \frac{a K_1 \epsilon
    \left(5 a^2+8 a (b+1)-4 (b-4) b\right) }{\kappa r^{2a + 3} } \label{504} \\
  p &= \frac{K_1 \epsilon  (a+2 b)
    \left(3 a^2+a (6 b+8)+8 b\right) }{\kappa r^{2 a+3}}.  \label{505} 
\end{align}
The sound speed has the constant value 
\begin{align}
\frac{dp}{d\rho} &= \frac{3 a^2+a (6 b+8)+8 b}{a (5 a-2 b+8)} \label{506} 
\end{align}
while the expression for the energy conditions are given by
\begin{align}
  \rho - p &= \frac{2 K_1 \epsilon
    \left(a^3-2 a^2 b-4 a b (2 b+1)-8 b^2\right) }{\kappa r^{2 a+3} } \\
  \rho + p &= \frac{4 (a+2) K_1 \epsilon  \left(2 a^2+5 a b+2 b^2\right) }{\kappa r^{2 a+3} } \\
  \rho + 3p &= \frac{2 K_1 \epsilon
    \left(7 a^3+2 a^2 (11 b+8)+4 a b (4 b+11)+24 b^2\right) }{\kappa r^{2 a+3} } 
\end{align}
and lastly, Chandrasekhar’s adiabatic index (also a constant in this case) is given by
\begin{align}
  \left(\frac{\rho +p}{p} \right)\frac{dp}{d\rho} &=& \frac{4 (a+2) (2 a+b)}{a (5 a-2 b+8)}
\end{align}
where $K_1 $ is a constant and  it must be recalled that $b = 2a+4 \pm \sqrt{7a^2 + 24a + 16}$. 
For the density to remain regular at the centre of the distribution demands
\[
a < -\frac{3}{2}
\]
leading to polynomial solutions. 

Given that the sound speed $\frac{dp}{d\rho}$ and adiabatic stability index are essentially constant depending only on $a$ and $b$ and not $r$, we may ask: what is the condition on $a$  to have $0< \frac{dp}{d\rho} < 1$ and $\left(\frac{\rho + p}{p}  \right)\frac{dp}{d\rho} > \frac{4}{3}$ satisfied simultaneously? Introducing (\ref{ab}) into the sound speed and adiabatic stability we get, respectively, 
\begin{align}
  \frac{dp}{d\rho} &=
  -\frac{8 \left(\sqrt{a (7 a+24)+16}+4\right)+
    a \left(11 a+4 \sqrt{a (7 a+24)+16}+40\right)}{a (3 a+8)} \label{321} \\[2ex]
  \left(\frac{\rho +p}{p} \right)\frac{dp}{d\rho} &=
  -\frac{4 (a+2)}{-2 a+\sqrt{a (7 a+24)+16}-4}
\end{align}
for the positive branch of (\ref{ab}). Now demanding causal behaviour as well as adiabatic stability gives the value of $a$ in the range
\begin{align}
\frac{1}{3} \left(-\sqrt{7}-5\right)<a<\frac{4}{7} \left(-\sqrt{2}-3\right)
\end{align}
or $-2.54858<a<-2.52241$ in decimal form. Note how thin this window is.  Curiously this interval happens to satisfy the requirement for a nonsingular centre namely $a < -\frac{3}{2}$. If we took the negative branch of (\ref{ab}) then the condition on $a$ works out to $-0,90616 < a < -0.78475$ which violates the regularity condition $a < -\frac{3}{2}$ so we omit it from further consideration. 

Unfortunately, the solution above is not suitable to model closed compact objects. Avoiding the singularity at the centre of the distribution by choosing $a < -\frac{3}{2}$ results in the density and pressure both going as $\displaystyle{\frac{1}{r^{2a+3}}}$ so that there will not be a pressure-free boundary in existence. In any case both coefficients of $\displaystyle{\frac{1}{r^{2a+3}}}$ will have to be positive for realistic behaviour so that both density and pressure will be monotonically increasing power-law functions forever from the centre. This is not physically acceptable. While we were able to detect exact solutions, it turns out that these solutions are not physically viable. 

\section{Potentials related proportionally} 

We again use the substitutions $\mu' = u$, $\nu' = v$, $\mu'' = u'$ and $\nu'' = v'$ that transform the isotropy equation to the first order form 
\begin{equation}
\label{eq:Q1}
\begin{aligned}
0 &= 4 r\,u'v^{2}
 - 2\,u^{3}\big(5 rv+1\big)
 - 3 ru^{4} 
 + u^{2}\left(6 r\big(u'+v'\big)-7 rv^{2}-6v\right) \\
&\quad
 + 2u\,v\left(6 ru'+4 rv'+ rv^{2}-2v\right).
\end{aligned}
\end{equation}

Let us  investigate a proportionality ansatz.  
Setting $v=au$ with constant $a$, equation \eqref{eq:Q1} reduces to
\begin{equation}
\label{eq:reduced}
u^2\,(2a+1)\Big[(a+1)\big(6 ru'-2u\big)
+ ru^2\big(a^2-4a-3\big)\Big] \;=\; 0.
\end{equation}
Hence either $u\equiv 0$ (and $v\equiv 0$), or provided $a\neq -1, -\frac{1}{2}$—$u$ solves the first-order ODE
\begin{equation}
\label{eq:firstorder}
6 r\,u'-2\,u+\frac{a^2-4a-3}{\,a+1\,}\,r\,u^2 \;=\; 0.
\end{equation}
equivalently 
\begin{equation}
6 r\,u'-2\,u+A\,r\,u^2=0, 
\qquad
A =\frac{a^2-4a-3}{a+1}, \label{531}
\quad a\neq -1.
\end{equation}
Equation \eqref{531} is a first-order Riccati type equation in $u$ however more easily recognisable  as a logistic equation. As Riccati equations are not solvable in general, we describe a method to solve equation (\ref{531}). Divide (\ref{531}) by $6r$ then 
\begin{align}
u'-\frac{1}{3r}\,u+\frac{A}{6}\,u^2=0 \label{532}
\end{align}
which is of Bernoulli type. Now let $y(r)=\frac{1}{u}$ so $y'(r)=-\frac{u'}{u^2}$. Then (\ref{532}) assumes the simple form 
\[
y'(r)+\frac{1}{3r}\,y(r)=\frac{A}{6}.
\]
as a first order linear differential equation. Using the integrating factor method with $\mu(r)=e^{\int (1/3r)\,dr}=r^{1/3}$ we get,
\[
\frac{d}{dr}\big(r^{1/3}y(r)\big)=\frac{A}{6}\,r^{1/3}
\;\Longrightarrow\;
r^{1/3}y(r)=\frac{A}{6}\int r^{1/3}dr + C
=\frac{A}{8}\,r^{4/3}+C.
\]
Hence
\begin{align}
u(r)=\frac{1}{y(r)}=\frac{r^{1/3}}{\,C+\dfrac{A}{8}\,r^{4/3}}\!, 
\qquad
v(r)=a\,u(r). \label{533}
\end{align}
For $a\neq -1$, \eqref{eq:firstorder} integrates to
\begin{align}
\mu = \frac{6}{A} \log \left(A r^{4/3}+8 C\right) \label{Aneq0}
\end{align}
provided $A \neq 0$. 
The dynamical quantities are calculated as follows. To keep the expressions transparent we introduce the radial variable $x\equiv r^{1/3}$, so that $r_1\equiv Ar^{4/3}+c_2=Ax^{4}+c_2$ and every fractional power reduces to a polynomial in $x$. It is further convenient to abbreviate the recurring parameter combinations
\begin{align}
\alpha_1=a(A+6)+A+3,\qquad \alpha_2=a(A+2)+A+3,\qquad \alpha_3=2a(A+3)+2A+3, \label{alphadef}
\end{align}
which depend on $(a,A)$ alone. The density and pressure then read
\begin{align}
\rho = -\frac{4096\,\epsilon\,x\,\Pi_\rho(x)}{k\,r_1^{\frac{12}{A}+5}}, \qquad
p = \frac{512(2a+1)\,\epsilon\,\Pi_p(x)}{k\,r_1^{\frac{4(A+3)}{A}}}, \label{eq:prop_rhop}
\end{align}
where the numerator polynomials are abbreviated as
\begin{align}
\Pi_\rho(x)=x\big[(2a(A+4)+A+5)x^{3}-8a^{2}\big]-3(2a+1)c_2, \qquad
\Pi_p(x)=\alpha_1 x^{4}+(a+1)c_2. \label{eq:prop_Pi}
\end{align}
The sound speed is found to be
\begin{align}
\frac{dp}{d\rho}=-\frac{3(2a+1)\,x^{3}\,r_1\,\mathcal{N}(x)}{\mathcal{D}(x)}, \qquad
\mathcal{N}(x)=c_2\alpha_2+(A+4)\alpha_1 x^{4}, \label{eq:prop_ss}
\end{align}
where, with all powers written as polynomials in $x$, the denominator is
\begin{align}
\mathcal{D}(x)=6x^{5}\big[\alpha_4 x^{3}-16a^{2}(3A+8)\big]
-2c_2\,x\big[\alpha_5 x^{3}-16a^{2}\big]+6(2a+1)c_2^{2}\label{eq:prop_D}
\end{align}
and we have additionally substituted $\alpha_4 = 5A^{2}+41A+80+2a(5A^{2}+36A+64)$ and $\alpha_5 = 62A+169+4a(31A+82.$ 
Since $\rho$ and $p$ in \eqref{eq:prop_rhop} carry the common factor $r_1^{-(12/A+5)}$ once the pressure is written over $r_1=Ax^{4}+c_2$, the energy conditions reduce to the compact closed forms
\begin{align}
\rho+p &= \frac{512\,\epsilon}{k\,r_1^{\frac{12}{A}+5}}\Big[(2a+1)\,\Pi_p(x)\,r_1-8x\,\Pi_\rho(x)\Big], \label{eq:prop_rpp} \\[1ex] 
\rho-p &= -\frac{512\,\epsilon}{k\,r_1^{\frac{12}{A}+5}}\Big[(2a+1)\,\Pi_p(x)\,r_1+8x\,\Pi_\rho(x)\Big], \label{eq:prop_rmp} \\[1ex]
\rho+3p &= \frac{512\,\epsilon}{k\,r_1^{\frac{12}{A}+5}}\Big[3(2a+1)\,\Pi_p(x)\,r_1-8x\,\Pi_\rho(x)\Big]. \label{eq:prop_rp3}
\end{align}
The Chandrasekhar adiabatic stability index then takes the form
\begin{align}
\Gamma=\left(\frac{\rho+p}{p}\right)\frac{dp}{d\rho}
=-\frac{3x^{3}\,\mathcal{N}(x)\,\big[(2a+1)\,\Pi_p(x)\,r_1-8x\,\Pi_\rho(x)\big]}{\Pi_p(x)\,\mathcal{D}(x)}. \label{eq:prop_Gamma}
\end{align}
The bracket in the numerator of \eqref{eq:prop_Gamma} is precisely the $\rho+p$ of \eqref{eq:prop_rpp}, so that causality and the stability bound are controlled by the single polynomial $\mathcal{D}(x)$ together with $\Pi_\rho$ and $\Pi_p$. We note in passing that the coefficient $2a^{2}+3a+1=(2a+1)(a+1)$ pervading these combinations vanishes exactly at $a=-\tfrac12$ and $a=-1$, the two values excluded in the reduction \eqref{eq:reduced}, so the causal and stable branch is automatically bounded away from these degeneracies.

To check if the solution from the assumption $v = a u$ results in physically reasonable star models, we now build a full star model by picking appropriate values for the free parameters. We analyze how density, pressure, sound speed, adiabatic stability and energy conditions change from the center to the surface of the star. This helps us check if the model is regular, obeys causality, is stable and is physically acceptable.

\begin{figure}[!htb]
\centering

\begin{subfigure}[t]{0.48\linewidth}
  \centering
  \includegraphics[width=\linewidth]{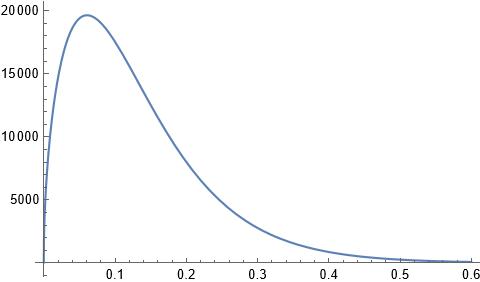}
  \caption{The density $\rho(r)$ for the proportional potential model.}
  \label{fig:rho}
\end{subfigure}\hfill
\begin{subfigure}[t]{0.48\linewidth}
  \centering
  \includegraphics[width=\linewidth]{pressure.jpg}
  \caption{Pressure $p(r)$ for the proportional potential model.}
  \label{fig:pressure}
\end{subfigure}

\vspace{0.8em}

\begin{subfigure}[t]{0.48\linewidth}
  \centering
  \includegraphics[width=\linewidth]{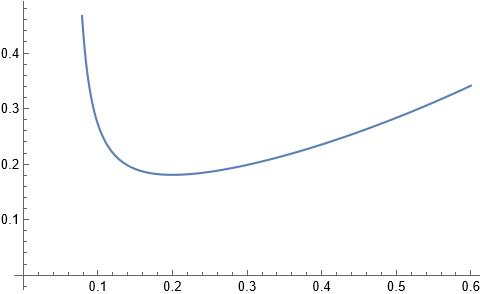}
  \caption{The behaviour of the sound speed $dp/d\rho$ for the proportional potential model.}
  \label{fig:SS}
\end{subfigure}\hfill
\begin{subfigure}[t]{0.48\linewidth}
  \centering
  \includegraphics[width=\linewidth]{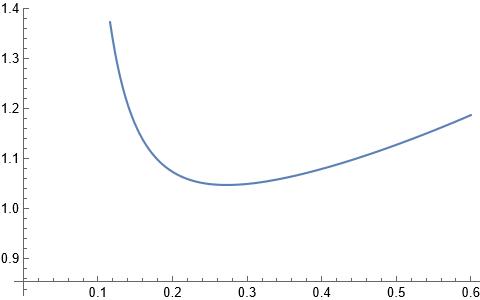}
  \caption{Adiabatic stability $\Gamma=\left(\frac{\rho+p}{p}\right)\frac{dp}{d\rho}$ for the proportional potential model.}
  \label{fig:AD}
\end{subfigure}

\vspace{0.8em}

\begin{subfigure}[t]{0.48\linewidth}
  \centering
  \includegraphics[width=\linewidth]{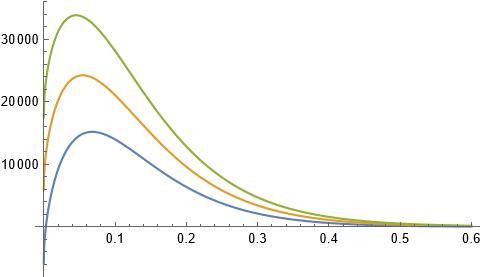}
  \caption{Energy conditions for the proportional potential model.}
  \label{fig:EC}
\end{subfigure}

\caption{Physical analysis plots for the proportional potential model.}
\label{fig:prop_potential_plots}
\end{figure}

Figure 2(a) shows the behaviour of the density $\rho(r)$. The density is positive, increase for a while and decreases monotonically with increasing radius and no singularities that are appearing. Figure 2(b) shows the behaviour of the pressure $p(r)$. From the expression in Eq.~\eqref{eq:prop_rhop}, the boundary condition $p(r_b)=0$ yields
\[
r_b^{4/3}\big(a (A+6)+A+3\big)+(a+1)c_2=0.
\]
Hence, the radial pressure vanishes at a finite boundary radius, which defines the surface of the compact object. In the special case $a=-\tfrac12$, one has $(2a+1)=0$ and therefore $p(r)\equiv0$.
 The sound speed $\frac{dp}{d\rho}$ is displayed in Figure 2(c). Although the sound speed remains within the causal bound between 0 and 1, it has a divergence at the centre. This singularity weakens the physical reliability of the model. Figure 2(d) shows the adiabatic stability index dropping to about 1.1 on the plot.  In stars the adiabatic stability is required to be greater than 4/3  in general relativity. This behaviour may be understood to mean that the stability index formula may need to be modified for $f(G, B)$ gravity.  Finally, the energy conditions are shown in Figure 2(e). All the energy conditions (weak, strong and dominant) remain positive throughout the stellar interior for the most part. We do observe a dip below zero very close to the centre where we have found a singularity exists anyway. This means the model is not well behaved everywhere. 
 
 To accommodate the misbehaving centre it is customary to remove a sphere of small radius at the centre and substitute it with another well behaved fluid. In other words our model will now have multiple layers - a behaviour not uncommon in realistic astrophysical objects. Our well behaved solution will now need to be matched with the regular centre across a common boundary. Given that there will be two exact solutions at play besides the vacuum outside, there will be a total of 4 necessary integration constants arising from second order differential equations. The outer solution will settle two of these constants using the vacuum and expressing the constants in terms of the mass and radius of the star. Now that these outer constants are solved they can be used to determine the inner sphere constants through matching. In this way the model may be completely determined. We can conclude that our fluid construction corresponds to a realistic shell of perfect fluid. 

It is worth situating these solutions against observational benchmarks. The proportional-potential shell, once its misbehaving core is excised and replaced by a regular interior, is closed by matching across the inner boundary and to the curved vacuum \eqref{vac2} at the surface; the two exterior constants are thereby fixed in terms of the gravitational mass $M$ and the radius $r_b$, so that a mass-radius relation follows numerically once the core solution is prescribed. A systematic confrontation with pulsar mass-radius data --- for example the $\sim 2\,M_\odot$ constraint from PSR~J0740$+$6620 or the NICER radius inferences --- would proceed along exactly these lines and is a natural next step. The isothermal branch to be considered in the next section, by contrast, carries a $1/r^2$ density profile by design and no finite surface; its natural observational analogue is therefore not a compact star but the flat rotation curves of galactic haloes \cite{BinneyTremaine,Rubin}, precisely as in the Einstein case. A signature distinguishing the quadratic bulk theory appears in the power-law branches, where the sound speed and the adiabatic index reduce to radius-independent constants fixed entirely by the exponents in \eqref{506} --- and hence by $(a,b)$ --- a rigidity that sharply narrows the simultaneously causal and stable window and is, in principle, testable through the admissible compactness range.

For convenience the outcomes of the four ans\"atze are consolidated in Table~\ref{tab:summary}.

\begin{table}[H]
\centering

\renewcommand{\arraystretch}{1.3}
\begin{tabular}{@{}lccccl@{}}
\hline
Model (ansatz) & Regular & Finite & $0<\frac{dp}{d\rho}<1$ & $\Gamma>\frac43$ & Viability\\
 & centre & surface & & & \\
\hline
$\mu=$ const (spatial)              & --  & --  & --  & --  & vacuum / defective \\
$\nu=$ const (temporal)             & yes & yes & no  & no  & $\rho<0$; rejected \\
Inverse radius $u,v\propto 1/r$     & yes$^{\dagger}$ & no  & yes & yes & no surface; rejected \\
Proportional $v=au$                 & no  & yes & yes$^{\ddagger}$ & below $\frac43$ & viable \emph{shell} \\
Isothermal $\rho,p\propto 1/r^{2}$  & no  & no  & yes & yes & halo / universe \\
\hline
\end{tabular}

\caption{Summary of the physical viability of the ans\"atze considered. $^{\dagger}$Regular only for $a<-\tfrac32$; $^{\ddagger}$causal but with a central divergence in $dp/d\rho$.}
\label{tab:summary}
\end{table}

\subsection{ The special case $A = 0$}

From (\ref{Aneq0}) it is clear that the case $A=0$ must be treated separately and does not belong to the previous discussion. Setting $A = 0$ gives   $a=2\pm\sqrt{7}$. Then  (\ref{533}) reduces to the form 
\begin{equation}
u'-\frac{1}{3r}\,u =0. \label{534}
\end{equation}
The solution of (\ref{534}) gives the power law forms
\begin{equation}
   u(r)=C\,r^{1/3}  \qquad
   v(r)=a\,C\,r^{1/3}
\end{equation}
hence we get the solutions
\begin{equation}
\mu = \frac{3}{4} Cr^{4/3} + C_1 \qquad \nu = \frac{3}{4} aC r^{4/3} + C_2
\end{equation}
where $C_1, C_2$ are integration constants. 

Letting $N= e^{-\frac{3}{2} C} $ and again writing $x\equiv r^{1/3}$, the density and pressure evaluate to
\begin{align}
\rho = -\frac{c^{4}\epsilon N\,x}{8k}\,\Pi_\rho^{0}(x), \qquad
p = \frac{(2a+1)c^{3}\epsilon N}{8k}\,\Pi_p^{0}(x), \label{eq:A0_rhop}
\end{align}
with the numerator polynomials
\begin{align}
\Pi_\rho^{0}(x)=(8a+5)c\,x^{4}-8a^{2}c\,x-24(2a+1), \qquad
\Pi_p^{0}(x)=3(2a+1)c\,x^{4}+8(a+1). \label{eq:A0_Pi}
\end{align}
The sound speed evaluates to
\begin{align}
\frac{dp}{d\rho}=\frac{6(2a+1)\,x^{3}\,\mathcal{N}_0(x)}{\mathcal{D}_0(x)}, \qquad
\mathcal{N}_0(x)=3(2a+1)c\,x^{4}+2(2a+3), \label{eq:A0_ss}
\end{align}
with the polynomial denominator
\begin{align}
\mathcal{D}_0(x)=16a^{2}c\,x\big(3c\,x^{4}-1\big)-8a\big(6c^{2}x^{8}-41c\,x^{4}+6\big)-30c^{2}x^{8}+169c\,x^{4}-24. \label{eq:A0_D}
\end{align}
Since here $\rho$ and $p$ share the common factor $c^{3}\epsilon N/8k$, the energy conditions follow in closed form as
\begin{align}
\rho+p &= \frac{c^{3}\epsilon N}{8k}\Big[(2a+1)\,\Pi_p^{0}(x)-c\,x\,\Pi_\rho^{0}(x)\Big], \label{eq:A0_rpp} \\[1ex]
\rho-p &= -\frac{c^{3}\epsilon N}{8k}\Big[(2a+1)\,\Pi_p^{0}(x)+c\,x\,\Pi_\rho^{0}(x)\Big], \label{eq:A0_rmp} \\[1ex]
\rho+3p &= \frac{c^{3}\epsilon N}{8k}\Big[3(2a+1)\,\Pi_p^{0}(x)-c\,x\,\Pi_\rho^{0}(x)\Big], \label{eq:A0_rp3}
\end{align}
and the Chandrasekhar adiabatic index reads
\begin{align}
\Gamma=\left(\frac{\rho+p}{p}\right)\frac{dp}{d\rho}
=\frac{6x^{3}\,\mathcal{N}_0(x)\,\big[(2a+1)\,\Pi_p^{0}(x)-c\,x\,\Pi_\rho^{0}(x)\big]}{\Pi_p^{0}(x)\,\mathcal{D}_0(x)}. \label{eq:A0_Gamma}
\end{align}

The discussion of the physical properties of this model is omitted for the sake of avoiding repetition. Generally our finding was that the model failed to display the expected features after numerous speculative attempts on the constants. 

\section{Isothermal behaviour}

Isothermal fluids are a type of matter where the temperature remains constant throughout the distribution. In astrophysics, assuming a fluid is isothermal demands the density and pressure to be expressed in simple power-law forms such as $\rho(r) = \frac{A}{r^2}$ and $p_r(r) = \frac{B}{r^2}$. In other words they both obey an inverse square law. These forms simplify the field equations, making it possible to find exact solutions that would otherwise be very difficult to obtain. Studying isothermal fluids helps us understand how a matter distribution with uniform temperature behaves under gravity and provides a useful experiment for checking the physical viability of modified gravity theories. Moreover, isothermal solutions often serve as benchmarks to compare more complex models.

The isothermal sphere  is often used to model galaxies and dark-matter \cite{BinneyTremaine} as it  produces flat rotation curves that match observations. Such models help explain the distribution of matter in the universe and the formation of large-scale structures \cite{Rubin,Peebles}. In modified gravity such as $f(G,B)$, studying isothermal fluids is particularly useful because it allows one to explore how bulk terms affect matter distributions. By applying the isothermal assumption in these theories, one can analyze whether the resulting stellar structure remain physically reasonable and investigate deviations from predictions of general relativity. Overall, isothermal fluid models are valuable because they provide simple yet realistic descriptions of matter, offering insights into both stellar and galactic systems while serving as a useful tool for testing alternative theories of gravity \cite{Clifton}.

We assume exact isothermal profiles for all $r>0$ of the form
\[
\rho(r)=\frac{A}{r^{2}},\qquad p(r)=\frac{B}{r^{2}}
\]
with $A,B$ constants and insert into equations (\ref{106a}) and (\ref{106b}). Introducing the logarithmic radial variable $x=\ln r$ and writing  overdots for
$\frac{d}{dx}$ we get 
\[
\mu'=\frac{\dot\mu}{r},\quad \mu''=\frac{\ddot\mu-\dot\mu}{r^{2}},\qquad
\nu'=\frac{\dot\nu}{r},\quad \nu''=\frac{\ddot\nu-\dot\nu}{r^{2}},\qquad
e^{-2\mu}=e^{-2\mu(x)}.
\]
After dividing by $8\kappa r$ on both sides a direct substitution of these relations into the right-hand sides of the field equations
(\ref{106a}) – (\ref{106b}) shows that each RHS factorises as
\[
\text{RHS}=\frac{e^{-2\mu(x)}}{r^{3}}\;\mathcal{P}\!\big(\dot\mu,\ddot\mu,\dot\nu,\ddot\nu\big),
\]

The density reduces to
\begin{equation}
\rho
=
\frac{e^{-2\mu(x)}\,\varepsilon}{8\kappa\, r^{4}}
\,\dot\mu\,
\mathcal{P}_{\rho}
(\dot\mu,\ddot\mu,\dot\nu,\ddot\nu),
\end{equation}
where
\begin{align}
\mathcal{P}_{\rho} &=
-32(\ddot\mu-\dot\mu)\dot\nu
+8\dot\mu^{2}(\dot\nu-2)
+5\dot\mu^{3}
-4\dot\mu\Big[
6(\ddot\mu-\dot\mu)
+4(\ddot\nu-\dot\nu)
+\dot\nu^{2}
+8\dot\nu
\Big].
\end{align}

Similarly, the pressure is obtained as
\begin{equation}
p
=
\frac{e^{-2\mu(x)}\,\varepsilon}{8\kappa\, r^{4}}
\,
\dot\mu(\dot\mu+2\dot\nu)
\Big[
\dot\mu(3\dot\mu+6\dot\nu+8)
+8\dot\nu
\Big]
\end{equation}

where $\mathcal{P}$ is a polynomial (with constant coefficients) in the $x$-derivatives of
$\mu,\nu$. 
Both field equations take the form
\[
8\kappa\,r\rho=\frac{e^{-2\mu(x)}\,\varepsilon}{r^{3}}\,\mathcal{F}_\rho(\dot\mu,\ddot\mu,\dot\nu,\ddot\nu),
\qquad
8\kappa\,r p=\frac{e^{-2\mu(x)}\,\varepsilon}{r^{3}}\,\mathcal{F}_p(\dot\mu,\dot\nu),
\]
where $\mathcal{F}_\rho$ and $\mathcal{F}_p$ are explicit polynomials in the $x$-derivatives of the metric potentials.

On the other hand the left-hand sides satisfy
\[
8\kappa\,r\rho=\frac{8\kappa A}{r},\qquad 8\kappa\,r p_r=\frac{8\kappa B}{r},
\]
so both LHS scale as $\displaystyle{\frac{1}{r}}$.

For the radial parts to match for \emph{all} $r$, we must therefore have
\[
\frac{e^{-2\mu(x)}}{r^{3}}\sim \frac{1}{r}
\quad\Longrightarrow\quad
e^{-2\mu(x)}\sim r^{2}
\quad\Longrightarrow\quad
\ \mu(x)=-x+\text{const}.\ 
\]
Hence $\dot\mu\equiv -1$ and $\ddot\mu\equiv 0$. With $\mu$ now determined, the $r$-dependence has been completely found; the only  $x$-dependence on the RHS’s arise through $\dot\nu(x)$ and $\ddot\nu(x)$ inside $\mathcal{P}$. However, the LHS’s are proportional to $\displaystyle{\frac{1}{r}}$ with constants $A,B$ and thus contain no $x$-dependence at all. It follows that 
$\mathcal{P}(\dot\nu,\ddot\nu)$ must be constant in $x$, which forces
\[
\ddot\nu=0\quad\Longrightarrow\quad \nu(x)=\beta x+\nu_{0}
\quad\Longleftrightarrow\quad
\ \nu(r)=\beta\ln r+\nu_{0}\ 
\]
where $\beta$ is a constant of integration. 

From these calculations we may conclude that the isothermality requirements  $\rho,p_r\propto \frac{1}{r^2}$ for all radial values forces the metric potentials to obey  the
logarithmic (power-law) family
\[
\ \mu(r)=-\ln r+\mu_{0},\qquad \nu(r)=\beta\ln r+\nu_{0}\ 
\]
up to the constants $(\mu_0,\nu_0,\beta)$. Any non-linear $x$-dependence (e.g.\ $\nu=\beta x+g(x)$
with $g'\neq 0$) would introduce $x$-dependence into $\mathcal{P}$ and thus spoil the exact $r^{-2}$ fall-off of $\rho$ and $p$. Consequently, the logarithmic (power-law) form is  unique under the assumption of exact isothermal profiles.

Let us now analyse this unique solution thoroughly. We take 
\[
\mu(r)=\alpha\ln r+\mu_0,\qquad \nu(r)=\beta\ln r+\nu_0,
\]
so that
\[
\mu'=\frac{\alpha}{r},\ \ \mu''=-\frac{\alpha}{r^2},\qquad
\nu'=\frac{\beta}{r},\ \ \nu''=-\frac{\beta}{r^2},\qquad
e^{-2\mu}=e^{-2\mu_0}\,r^{-2\alpha}.
\]
Substituting  into the right-hand sides of (\ref{106a}) and (\ref{106b})  gives 
\begin{align}
  8\kappa \frac{A}{r} &= e^{-2\mu_0}\,\varepsilon\,
\frac{\alpha^2\left(5\alpha^2+8\alpha\beta+8\alpha-4\beta^2+16\beta\right)}{r^{3+2\alpha}}, \label{570a} \\[1ex]
8\kappa \frac{B}{r} &= e^{-2\mu_0}\,\varepsilon\,
\frac{\alpha(\alpha+2\beta)\big(\alpha(3\alpha+6\beta+8)+8\beta\big)}{r^{3+2\alpha}}. \label{570b}
\end{align}
Since the left-hand sides scale as \(r^{-1}\), we must have 
\[
3+2\alpha = 1 \quad\Longrightarrow\quad \ \alpha=-1\ 
\]
for the matching. 
With \(\alpha=-1\), equations (\ref{570a}) and (\ref{570b})  reduce to
\begin{align}
8\kappa A\,e^{2\mu_0}+\varepsilon\,(2\beta - 3)(2\beta - 1)&=0, \label{E1}\\
8\kappa B\,e^{2\mu_0}+\varepsilon\,(2\beta-5)(2\beta-1)&=0. \label{E2}
\end{align}
It is convenient to introduce the definition $s=\frac{B}{A}$. Eliminating \(e^{2\mu_0}\) and $\varepsilon$ between \eqref{E1}–\eqref{E2} yields
\[
s\,(4\beta^2-8\beta+3)=(2\beta-1)(2\beta-5)=4\beta^2-12\beta+5,
\]
whose  solution is
\[
\ \beta=\frac{3s-5}{2(s-1)\ },\qquad s\neq 1.
\]
Note that  \(\beta=\tfrac12\), $\frac{3}{2}$ and $\frac{5}{2}$ forces 
\(e^{2\mu_0}=0\) from \eqref{E1}, which is not acceptable. Thus $\beta = \frac{1}{2}, \frac{3}{2},  \frac{5}{2}$ are discarded.

Solving \eqref{E1} (or \eqref{E2}) for \(e^{2\mu_0}\) gives
\[
\ e^{2\mu_0}=-\,\frac{\varepsilon}{8\kappa A}\,\big(4\beta^2-8\beta+3\big)
\;=\; -\,\frac{\varepsilon}{8\kappa B}\,(2\beta-5)(2\beta-1)\ ,
\]
which fixes \(\mu_0\) once \(A,B,\kappa,\varepsilon\) are chosen. The additive constant \(\nu_0\) is free
(a rescaling of \(t\)).

In conclusion, for given constants \(A, B, \kappa,\) and \(\varepsilon\) with \(B \neq A\), we define
\[
s = \frac{B}{A},
\]
and obtain
\[
\mu(r) = -\ln r + \mu_0, \qquad
\nu(r) = \beta \ln r + \nu_0,
\]
where
\[
\beta = \frac{3s - 5}{2(s - 1)}, \qquad
e^{2\mu_0} = -\frac{\varepsilon}{8\kappa A}\,(4\beta^2 - 8\beta + 3).
\]

These \((\mu,\nu)\) reproduce exactly $\rho = \frac{A}{r^{2}}$ and $p_r = \frac{B}{r^{2}}$.
 in (\ref{106a})-(\ref{106b}) and generate isothermal behaviour. There are no other cases possible for isothermality. 

Now, let us investigate  the physical properties of the exact isothermal solution we found. Because this model is designed so that both density $\rho(r)$ and pressure $p(r)$ are proportional to $r^{-2}$ for all $r>0$ a singularity at the centre is unavoidable. 

\begin{figure}[!htb]
\centering

\begin{subfigure}[t]{0.48\linewidth}
  \centering
  \includegraphics[width=\linewidth]{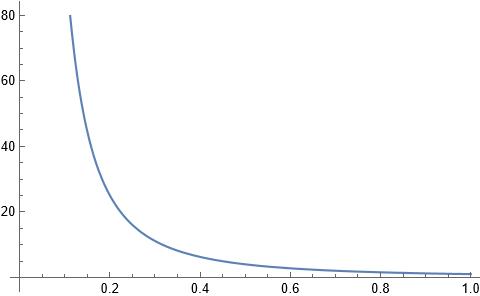}
  \caption{The density $\rho(r)=A/r^{2}$ for the isothermal model.}
  \label{fig:iso_rho}
\end{subfigure}\hfill
\begin{subfigure}[t]{0.48\linewidth}
  \centering
  \includegraphics[width=\linewidth]{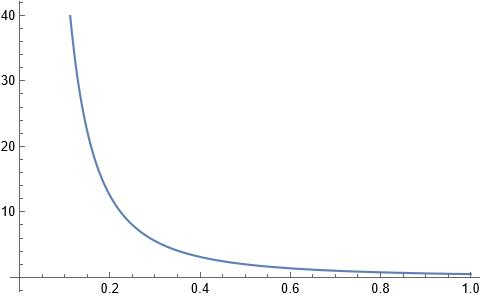}
  \caption{The pressure $p(r)=B/r^{2}$ for the isothermal model.}
  \label{fig:iso_pressure}
\end{subfigure}

\vspace{0.8em}

\begin{subfigure}[t]{0.48\linewidth}
  \centering
  \includegraphics[width=\linewidth]{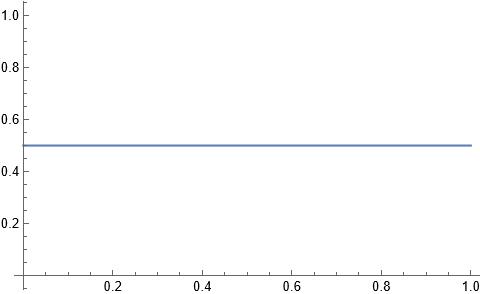}
  \caption{The sound speed $\frac{dp}{d\rho}$ for the isothermal model.}
  \label{fig:iso_SS}
\end{subfigure}\hfill
\begin{subfigure}[t]{0.48\linewidth}
  \centering
  \includegraphics[width=\linewidth]{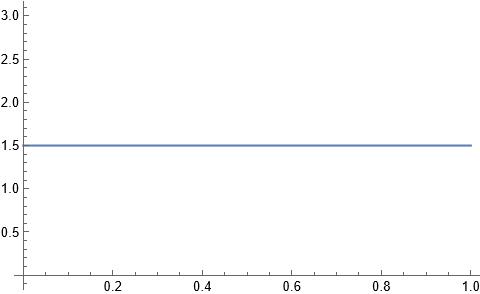}
  \caption{Adiabatic stability $\Gamma=\left(\frac{\rho+p}{p}\right)\frac{dp}{d\rho}$ for the isothermal model.}
  \label{fig:iso_AD}
\end{subfigure}

\vspace{0.8em}

\begin{subfigure}[t]{0.48\linewidth}
  \centering
  \includegraphics[width=\linewidth]{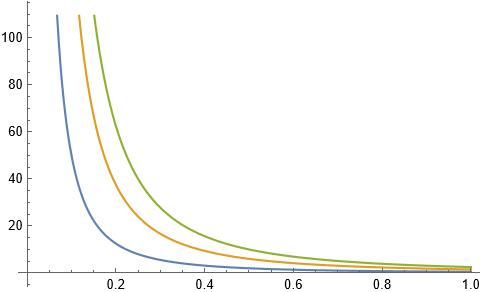}
  \caption{Energy conditions for the isothermal model.}
  \label{fig:iso_EC}
\end{subfigure}

\caption{Physical analysis plots for the isothermal model.}
\label{fig:isothermal_plots}
\end{figure}

\subsection{Physical analysis of the isothermal model}

We now study the physical behaviour of the exact isothermal solution obtained from the power law metric potentials.  For the  analysis, we choose the density constant $A=1$, the coupling parameter $\varepsilon=-1$ and $\kappa=1$. The sound speed is constant and is given by
\[
s \equiv \frac{dp}{d\rho} = 0.5 .
\]
The additive constant in the temporal potential is set to $\nu_0=0$. With these choices, the pressure constant becomes $B = sA = 0.5$.
Through systematic fine-tuning the remaining constants are fixed as $\beta=3.5$ and $e^{2\mu_0}=3$, which gives $\mu_0 = \tfrac{1}{2}\ln 3$.
These values fully determine the metric potentials and the scale of the solution.
The density is given by
\[
\rho(r)=\frac{1}{r^2},
\]
which is positive for all $r>0$. As $r$ approaches zero, the density becomes infinite, showing that the model has a central singularity. This behaviour is common in exact isothermal models \cite{saslawisothermal}. For all $r>0$, the density decreases with radius since
\[
\frac{d\rho}{dr} = -\frac{2A}{r^3} < 0.
\]
This means the matter becomes less dense as we move away from the centre.

The pressure has the same form,
\[
p(r)=\frac{B}{r^2} = \frac{sA}{r^2},
\]
and is positive everywhere because $s>0$. The pressure decreases with radius, but it does not become zero at any finite value of $r$. For example, at $r=1$ the pressure is $p=0.5$. The pressure only goes to zero as $r \to \infty$. Therefore, the solution does not describe a star with a fixed surface, but rather an infinitely extended isothermal matter distribution or a universe. The sound speed is constant and satisfies the causal condition $0 < \frac{dp}{d\rho} < 1$. The adiabatic index $\Gamma$ is also constant and has the value $\Gamma = 1.5$. Since $\Gamma > \tfrac{4}{3}$, the model is stable against small radial disturbances. This shows that, even though the model has a central singularity and no finite boundary, it is dynamically stable within this framework. Similar to the Einstein case, the isothermal behaviour in $f(G, B)$ gravity cannot describe a closed (bounded) compact object but rather a physically reasonable model of a universe.

\section{Conclusion}
  
We have endeavoured to generate exact astrophysical models in the bulk-boundary decomposition of the action as per BJ. The gravitational field equations are considerably more complicated than the Einstein equations and on writing the equation of pressure isotropy connecting the geometric quantities, it was found that the simplest choices for one of the two did not lead to tractable differential equations.  Consequently we decided on different tactics some of which were inspired by our earlier work \cite{hans-cgb-nb}.  We commenced with the simplest case of constant potentials. We found that these models did not give realistic stars. We then considered potentials varying inversely with radius. Complete exact models were possible however many physical constraints were compromised. Models in which the metric potentials were proportional to each other were more promising. They had regular central pressure and densities however they were singular at the origin. Nevertheless a positive feature was the presence of a finite boundary signifying closed compact distributions. Stability, causality and energy conditions were met in a suitable shell. Finally the isothermal solution with both density and pressure scaling as $\frac{1}{r^2}$ generated physically meaningful results. The isothermal model reproduces positive density, positive pressure and a monotonically decreasing density, while maintaining a constant sound speed. Furthermore, the adiabatic stability satisfies the $\Gamma> \frac{4}{3}$ bound.  The pressure does not vanish at the boundary, implying that the model represents an infinite sphere rather than a compact star with a well defined surface. The centre is obviously singular by design and the overall behaviour of the isothermal model mimics its Einstein counterpart in general relativity \cite{Dadhich}. While we were successful in generating some classes of exact solutions and showing that at least a few have pleasing properties, what is demonstrated in this work  is that the BJ proposition does have merit in the construction of stellar models. Our attempt represents the first such search for bounded distributions in this framework. 

A word on the status of the boundary term is in order. For the static, conformostatic configurations considered here the boundary contribution $B$ enters the action only as a total divergence and, in the pure-bulk theory built from $f(G)$ alone, does not source the field equations; the entire dynamical content resides in the bulk invariant $G$. This is exactly what preserves the second-order structure and distinguishes the construction from $f(R)$ gravity, in which the bulk and boundary parts are rigidly locked together. The situation changes in time-dependent settings: on a cosmological background $B$ is generically dynamical, so a genuine $f(G,B)$ treatment with both arguments active should yield a richer phenomenology and is a clear direction for future work. Further natural extensions include charged perfect fluids, the relaxation of isotropy to admit anisotropic stresses, slowly-rotating configurations, and a full multi-layer numerical construction of the shell model matched to a realistic equation of state.

\end{document}